\documentclass[a4paper,11pt]{article}
\usepackage{pos}

\title{Proton decay matrix elements on PACS configurations}

\author*[\dag, a]{Ryutaro Tsuji}
\author[b]{Yasumichi Aoki}
\author[c]{Yoshinobu Kuramashi}
\author[c]{Eigo Shintani}
\author{PACS Collaboration}

\affiliation[a]{High Energy Accelerator Research Organization (KEK), 305-0801, Tsukuba, Japan}

\affiliation[b]{RIKEN Center for Computational Science, 650-0047, Kobe, Japan}

\affiliation[c]{Center for Computational Sciences, University of Tsukuba, 305-8577, Tsukuba, Japan}

\emailAdd{rtsuji@post.kek.jp}

\abstract{
We report the preliminary results of lattice computation for the proton decay matrix elements in $N_f=2+1$ physical point with Wilson-clover fermion. We perform it on the PACS configurations of $64^4$ lattice volume with lattice spacing  $a = 0.085$ fm, and carefully estimate the systematic uncertainties, especially for the excited state contamination and associated error of the renormalization constant with Regularization Independent (RI, Rome-Southampton) scheme. Our preliminary results of the twelve relevant transition modes in proton decay matrix element and comparison with other lattice results are presented. 
}

\FullConference{The 41st International Symposium on Lattice Field Theory (LATTICE2024)\\
 28 July - 3 August 2024\\
Liverpool, UK\\}

\begin{document}
\maketitle
\section{Introduction}
\label{sec:introduction}
Proton decay is a smoking-gun signal of beyond the standard model (BSM) physics. The most recent experimental constraints of proton lifetime is \cite{Super-Kamiokande:2020wjk,JUNO:2022qgr,Super-Kamiokande:2022egr}
\begin{eqnarray}
    &&\tau(p\rightarrow \pi^0 e^+) > 2.4\times 10^{34}\,{\rm yrs.},\label{eq:sk_limit}\\
    &&\tau(p\rightarrow K^+\bar{\nu}) > 5.9\times 10^{33}\,{\rm yrs.},\label{eq:juno_limit}\\
    &&\tau(p\rightarrow K^0\mu^+) > 9.6\times 10^{33}\,{\rm yrs.},\label{eq:sk_limit2}
\end{eqnarray}
which are used for the limitation of parameters in the Grand Unified Theories (GUTs) and SUSY-GUTs. The next generation of underground neutrino detector, such as DUNE, JUNO and Hyper-Kamiokande (HK), will reach the high sensitivities at the level of $10^{34}$--$10^{35}$ years~\cite{Ellis:2019fwf} in which the GUT models predict the occurrence decaying the proton into meson and anti-lepton through, for instance, X particle exchange in GUTs and Higgsino exchange in SUSY-GUTs. Those new limitations which are an order of magnitude stronger than the current ones will also play a crucial role to expand an exclude region of many suggested (SUSY-)GUT models.

To evaluate a lifetime of proton from (SUSY-)GUTs, the "proton decay matrix elements", which is the transition form factor from proton to pseudoscalar meson through baryon number violating operator, are indispensable quantities, since the partial width of such a decay process is effectively decomposed into the Wilson coefficients depending on the (SUSY-)GUTs parameters in high energy scale and the proton decay matrix element determined from the SM in low energy scale~\cite{JLQCD:1999dld,Aoki:2006ib}. The matrix elements are nonberturbative quantities given from quark dynamics and lattice QCD (LQCD) then plays an essential role to evaluate it from the first principles of QCD.

In the most recent LQCD computation of the proton decay matrix elements with domain wall fermions in $N_f=2+1$ physical point ensembles~\cite{Yoo:2021gql}, the result without the chiral extrapolation, which had been the one of the large sources of systematic uncertainty in the previous study~\cite{Martin:2011nd,Aoki:2017puj}, was reported. They also reported the numerical study of systematic uncertainties of the excited state contamination, the renormalization constant and discretization error, and consequently their continuum results were 10--20\% precision. 

In this proceedings, we report our preliminary results for proton decay matrix elements in $N_f=2+1$ physical point ensembles with Wilson-clover fermion in $L=64$ lattice box with 2.3 GeV lattice cutoff scale. Here we especially update our numerical study of the renormalization constant from our previous report~\cite{Aoki:2020qus}, and the comparison with the recent computation~\cite{Yoo:2021gql} is also shown.

\section{Matrix elements of proton decay through the baryon number violation operator}
\label{sec:matrix_elements_of_proton_decay_through_the_baryon_number_violation_operator}
We can define the matrix elements of proton (also neutron) decay as the transition matrix elements from an initial nucleon state with momentum $k$ to the final pseudoscalar state with momentum $p$ via the baryon number violating three-quark operator $\mathcal O^I$, where label $I$ indicates a certain flavor structure, for instance $I=RL$ when $\mathcal O^I=\varepsilon_{ijk}(u^{i\,T}CP_Rd^j)P_Lu^k$ for $p\rightarrow\pi^0$ mode, read into
\begin{eqnarray}
    \langle P(p)|\mathcal O^{RL}|N(p,s)\rangle = P_L\Big[ W^{RL}_0(q^2)-\frac{iq\hspace{-1.8mm}/}{m_N}W_1^{RL}(q^2)\Big]u_N(k,s),\label{eq:w0,w1}
\end{eqnarray}
with the chiral projection matrix $P_R$ for right-handed and $P_L$ for left-handed ones. The transition form factors $W_0$ and $W_1$ are the functions of the squared momentum transfer $q^2$, defined as $q=k-p$, and those are regarded as the relevant and irrelevant form factors respectively to our target kinematics, which mean that with the on-shell lepton condition on $e^+$ and $\bar\nu$ and $\mu^+$ the second term in Eq.(\ref{eq:w0,w1}) should be irrelevant because of $m_l/m_N\simeq 0$ (for $l=\mu^+$ this will not be satisfied for more rigorous estimate, and strictly speaking we need to take into account $W_1$ contribution in this case \cite{Aoki:2017puj,Yoo:2021gql}). Consequently the partial decay width of proton decay, $p\rightarrow P+\bar l$, can be expressed as
\begin{eqnarray}
    \Gamma(p\rightarrow P+\bar l) = \frac{m_N}{32\pi}\left[1-\left(\frac{m_P}{m_N}\right)^2\right]\left|\sum_{I=LR,LL}C^IW_0^I(0)\right|^2,\label{eq:partial_width}
\end{eqnarray}
with the relevant form factors, $W_0^I$ in Eq.(\ref{eq:w0,w1}) at the physical kinematics, $-q^2=m_l^2\simeq 0$. In the above equation, the chiral-pair contribution can be reduced from four to two, $I=RL,LL$, which is as a consequence of the parity symmetry~\cite{Aoki:2006ib,Aoki:2007xm,Aoki:2017puj}. Here we treat the QCD contribution with LQCD at the low energy scale, $O(\Lambda_{\rm QCD})$, for the form factors, and the GUTs and SUSY-GUTs model parameters are only included in $C_I$ after down to the hadronic scale\cite{Ellis:2019fwf}. The LQCD can non-perturbatively provide the relevant form factor $W^I_0(0)$ in Eq.(\ref{eq:partial_width}), and it is then a fundamental element for the evaluation of partial decay width in Eq.(\ref{eq:partial_width}).

\section{Simulation detail}\label{sec:simulation_detail}

\begin{table}[t]
  \begin{center}
    \caption{Summary of the lattice parameters for the gauge field configurations, source smearing parameters and source-sink separation($t_{\rm sep}$), low-precision source points in AMA ($N_s^{\rm AMA}$) used in this work.}
    \label{tab:simulation_parameter}
    \begin{tabular}{ccccccc}
      \hline\hline      
       $L^3\times L$ & $a^{-1}$ (GeV) & $m_\pi$ (MeV) & Source smearing & $t_{\rm sep}$ & \#configs & $N_s^{\rm AMA}$ \\
      \hline
      $64^3\times 64$ & 2.3162(44) & 139 & Gauss. & 18 & 45 & 64\\
      & & & & 20 & 51 & 256\\
      & & & & 24 & 50 & 384\\
      & & 135 & Exp. & 20 & 396 & 32\\
      \hline\hline
    \end{tabular}
  \end{center}
\end{table}

We use the $N_f=2+1$ flavor physical point gauge configurations generated by the PACS Collaboration with the six stout-smeared ${\mathscr{O}}(a)$ improved Wilson-clover quark action and Iwasaki gauge action at $\beta=1.82$ corresponding to the lattice spacings of $0.09$ fm~\cite{Ishikawa:2018jee, Ishikawa:2021eut}. The lattice volume $64^4$ that has 5.4 fm for a spatial extent is used in this study. When we compute the hadron correlation functions, the all-mode-averaging (AMA) technique~\cite{Blum:2012uh,Shintani:2014vja} with the deflated Schwartz Alternative Procedure(SAP)~\cite{Luscher:2007se} for Wilson-clover fermion~\cite{vonHippel:2016wid} is employed in order to reduce the statistical errors significantly without increasing computational costs~\cite{Shintani:2018ozy, Tsuji:2022ric, Tsuji:2023llh}.  Table~\ref{tab:simulation_parameter} shows the simulation parameters including the number of configurations, source/sink smearing types, we employ both exponential with $(A,B)=(1.2,0.33)$ and Gaussian $(W,N)=(10,600)$ smearings
~\footnote{In this study, both of the exponentially smeared quark operator $q_S(t,\boldsymbol{x})=\sum_{\boldsymbol{y}}A\mathrm{e}^{-B|\boldsymbol{x} \boldsymbol{y}|}q(t,\boldsymbol{y})$ with the Coulomb gauge fixing and the gaussian smeared quark operator $q_G(t,\boldsymbol{x})=\sum_{\boldsymbol{y}}\left[\delta_{\boldsymbol{x},\boldsymbol{y}}+\frac{W^2}{4N}\boldsymbol{\Delta}\right]q(t,\boldsymbol{y})$ with the three-dimensional covariant Laplacian $\boldsymbol{\Delta}$ used for the construction of the interpolating operator as well as a local quark operator $q(t,\boldsymbol{x})$. For detail, see Ref.~\cite{Tsuji:2023llh} and references therein.}
, and the source-sink separation, $t_{\rm sep}/a=\{18, 20, 24\}$ for $m_\pi=139$ MeV and $t_{\rm sep}=20$ for $m_\pi=135$ MeV, for the three-point function, of which the sequential source method is employed for the quark line between the time-slices of operator and sink positions. $N_s^{\rm AMA}$ denotes the number of source points with low-precision quark propagator, which is obtained by that the number of iteration in the solver algorithm is constrained in a small number (since we use a mixed precision Generalized Conjugate Residual (GCR) algorithm, we choose it as the constrained outer-double-precision GCR iteration). Here the low-precision quark propagator is given by the solver with four GCR iterations. In addition, we tune the lattice momentum transfers onto the physical kinematics ($q^2=0$) as the interpolation point.

\section{Renormalization}
\label{sec:renormalization}
The renormalization constants of the baryon number violating three-quark operator have been conventionally given from $\overline{\rm MS}$ subtraction scheme at 2 GeV, $Z^{\overline{\rm MS}}_{O=LL,RL}(2\,{\rm GeV})$~\cite{JLQCD:1999dld,Aoki:2006ib,Aoki:2007xm,Yoo:2021gql}. In this work, we first evaluate the renormalization constants with the Rome-Southampton scheme~\cite{Martinelli:1994ty} using LQCD evaluation of quark propagator under the Landau gauge, and via the perturbative matching procedure we can then calculate the target renormalization constants. According to \cite{Yoo:2021gql}, we employ both of the $\mathrm{MOM_{3q}}$~\cite{Aoki:2006ib} scheme, where all three quarks carry the same momenta ($p=k=r$),  and $\mathrm{SYM_{3q}}$~\cite{Gracey:2012gx} scheme, where all three quarks carry momenta of the same magnitude ($p^2=k^2=r^2)$ that add to zero momentum ($p+k+r=0$). As for the wavefunction renormalization constant $Z_q$, we employ the $\mathrm{RI/SMOM}$ and $\mathrm{RI/SMOM}_{\gamma_\mu}$ schemes~\cite{Aoki:2007xm, Sturm:2009kb,Almeida:2010ns}. In LQCD evaluation of quark propagator, we use 100 configurations for $m_\pi=139$ MeV in Table~\ref{tab:simulation_parameter}.

\begin{figure*}
\centering
\includegraphics[width=0.49\textwidth,bb=0 0 702 612,clip]{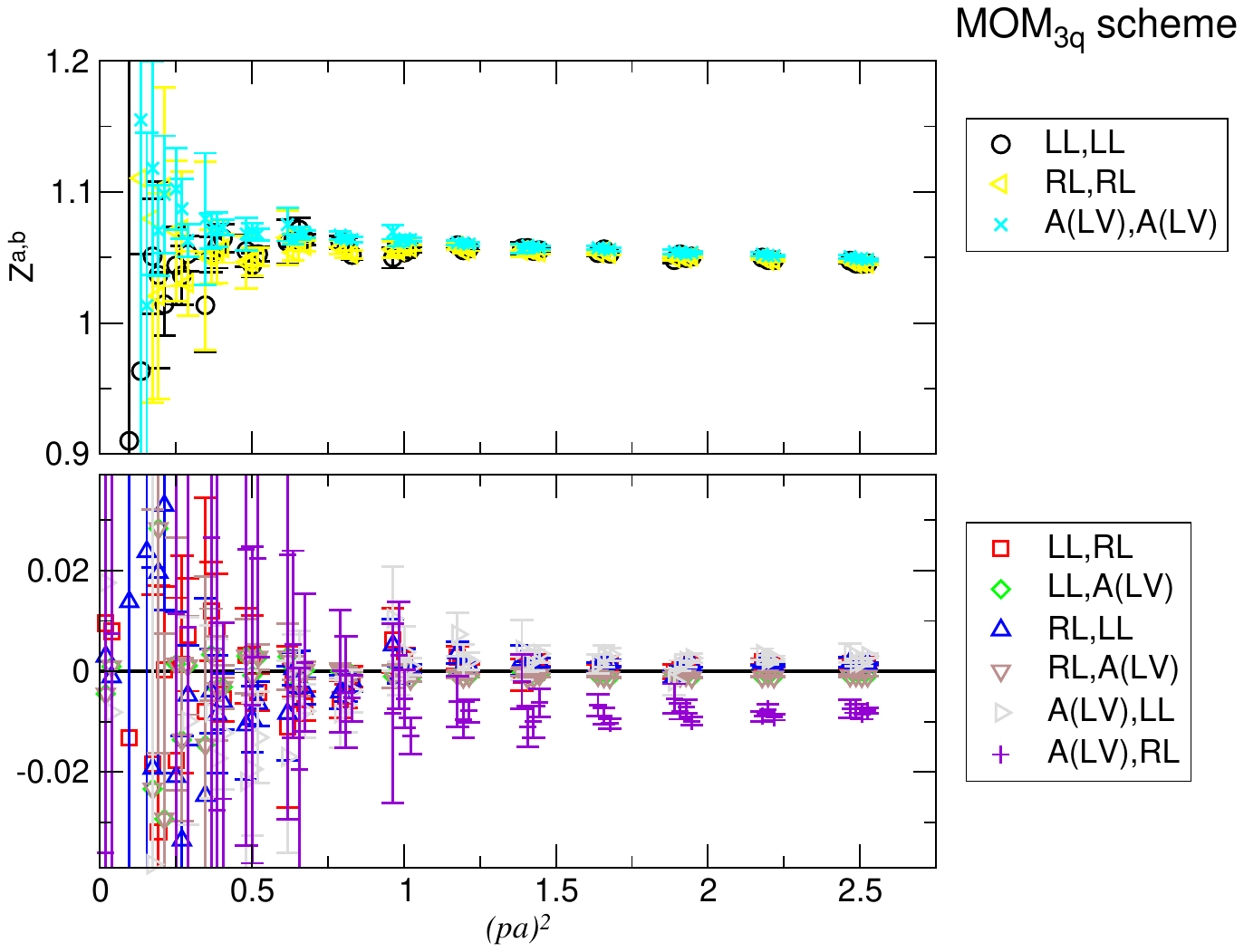}
\includegraphics[width=0.49\textwidth,bb=0 0 702 612,clip]{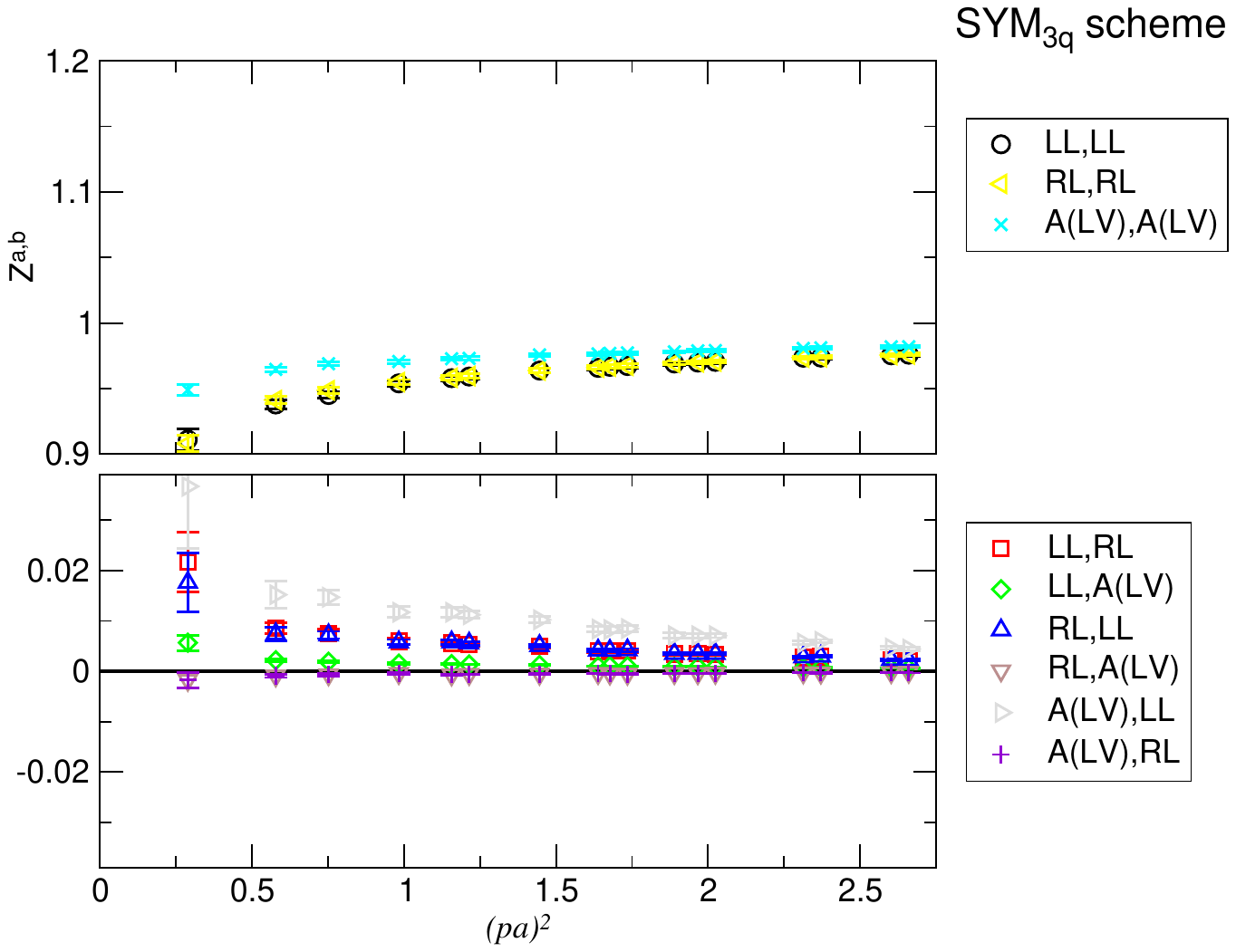}
\caption{
Elements of renormalization $3\times 3$ matrix with the $\mathrm{MOM_{3q}}$ (left) and the $\mathrm{SYM_{3q}}$ (right) as functions of momentum squared in lattice units.
}
\label{fig:operator_mixing}
\end{figure*}

Figure~\ref{fig:operator_mixing} shows the each element of the renormalization $3\times3$ matrix described in Ref.~\cite{Aoki:2006ib} with two different schemes, $\rm MOM_{3q}$ and $\rm SYM_{3q}$, as a function of injected momentum squared in the lattice units. One can see that the size of the off-diagonal elements (lower-panel) are at most 2\% compared to the size of the diagonal elements (upper panel) for both schemes. We also observe that the bare form factors corresponding to the diagonal and off-diagonal elements have the very similar values, and it turns out that the operator mixing effect due to the chiral symmetry violation in Wilson-clover fermion is not significant for our form factor calculation~\cite{JLQCD:1999dld}. Hence, in this study, the three-quark operators can be multiplicatively renormalized within our precision.

%
%  FIG.2.1
%
%
%
%  FIG.2.2
%
\begin{figure*}
\centering
\includegraphics[width=0.49\textwidth,bb=0 0 792 612,clip]{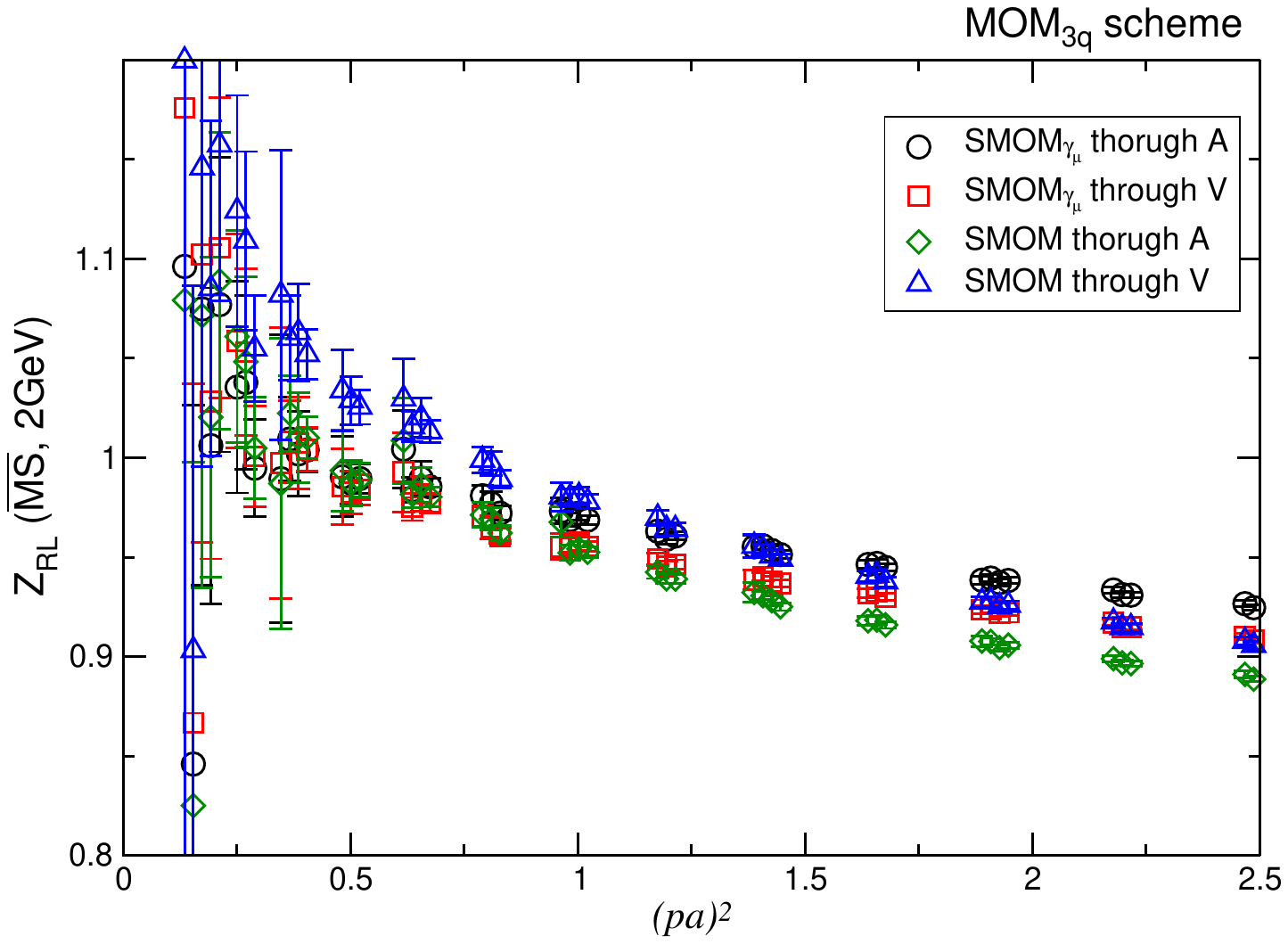}
\includegraphics[width=0.49\textwidth,bb=0 0 792 612,clip]{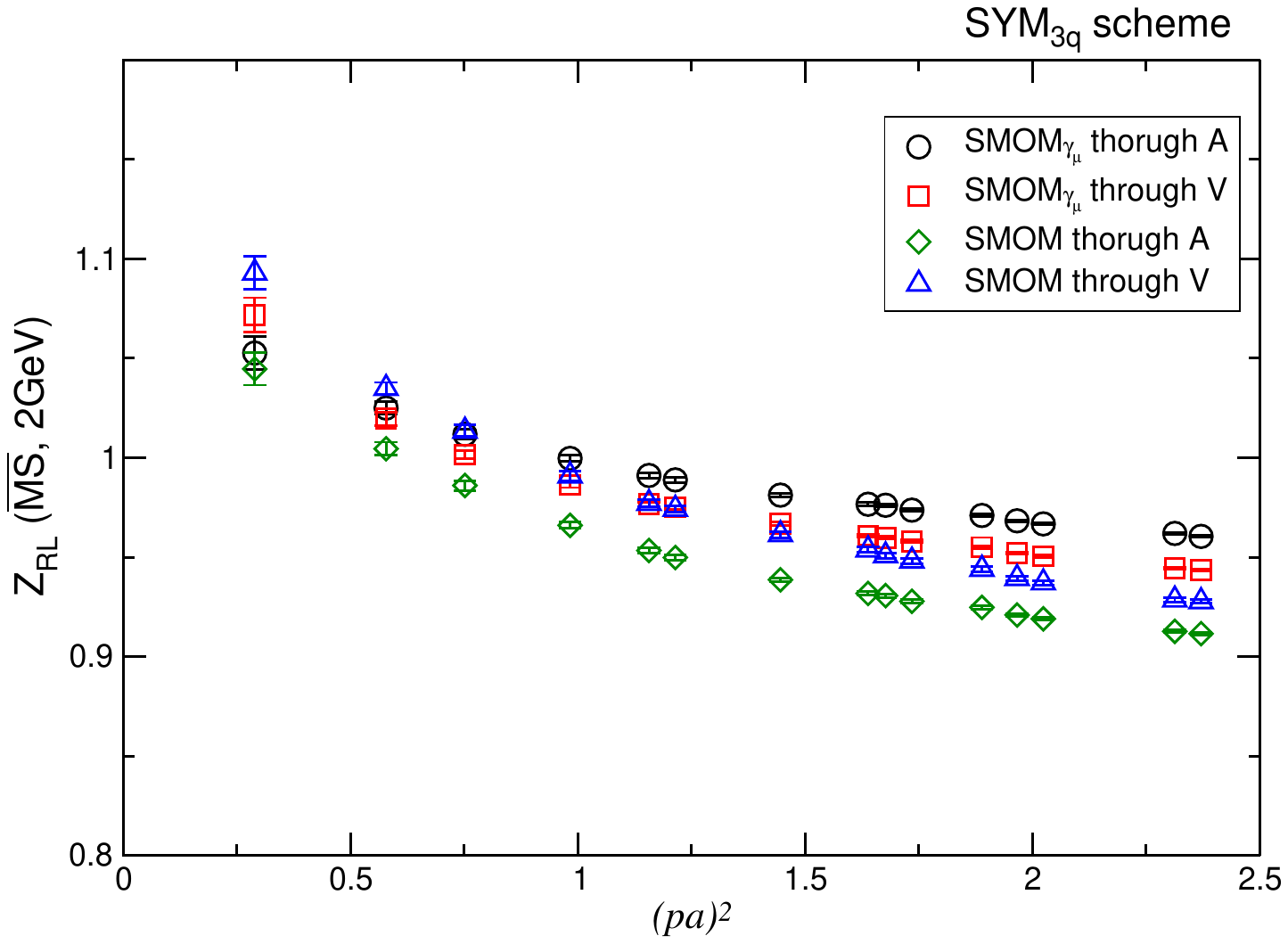}
\caption{
Renormalization constant in $\rm\overline{MS}$ scheme at 2 GeV with $O=RL$ via $\mathrm{MOM_{3q}}$ (left) and $\mathrm{SYM_{3q}}$ (right) with four different intermediate $\mathrm{RI/SMOM}_{(\gamma_\mu)}$ matching schemes (see, e.g. Ref~\cite{Tsuji:2022ric} and references therein.) as a function of matching scale squared.
}
\label{fig:renormalization_constant}
\end{figure*}

The renormalization constants obtained by $\rm MOM_{3q}$ and $\rm SYM_{3q}$ schemes are converted to the $\overline{\mathrm{MS}}$ scheme evolved to the scale of 2 GeV by the perturbation theory as
\begin{align}
    \label{eq:eq_matching}
    Z_O^{\mathrm{\overline{MS}}} (2\ \mathrm{GeV})
    =
    E_{O}(2\ \mathrm{GeV}, \mu_0)
    C^{x}_{O}(\mu_0)
    Z_O^{x}(\mu_0),\, x \in \{\mathrm{RI/SMOM},\mathrm{RI/SMOM}_{\gamma_\mu}\},
\end{align}
with the evolution factor $E_{O}(2 \mathrm{GeV}, \mu_0)=Z_O^{\mathrm{\overline{MS}}}(2\ \mathrm{GeV})/Z_O^{\mathrm{\overline{MS}}}(\mu_0)$ and conversion factor $C^x_{O}(\mu_0)=Z^{\mathrm{\overline{MS}}}_{O}(\mu_0)/Z^x_{O}(\mu_0)$. Figure~\ref{fig:renormalization_constant} shows the scale dependence of $Z_{O=RL}^{\mathrm{\overline{MS}}}$ converted from $\rm MOM_{3q}$ and $\rm SYM_{3q}$ schemes. There is the residual dependence on the choice of the matching scale $\mu_0$, and such a dependence is one of the systematic uncertainties in the determination of the renormalization constants. Our renormalization constants for $RL$ and $LL$ operators are evaluated as
\begin{align}
    Z_{RL}^{\overline{\mathrm{MS}}}(2\;\mbox
    {GeV}) 
    =
    1.016(5)_{\mathrm{stat}}(41)_{\mathrm{sys}}
    \ \mathrm{and}\ 
    Z_{LL}^{\overline{\mathrm{MS}}}(2\;\mbox
    {GeV}) 
    =  1.018(6)_{\mathrm{stat}}(37)_{\mathrm{sys}}
    ,
\end{align}
where the first error is statistical error, and the second one is the systematic uncertainty from the lattice artifacts, unwanted infrared divergence, uncertainties from perturbation and others, and intermediate scheme dependence (see Ref.~\cite{Tsuji:2022ric} in detail), in which the total systematic errors are evaluated as the root-mean-squared sum of those systematic uncertainties.

\section{Results of proton decay matrix elements}
\label{sec:renormalized_proton_decay_matrix_elements}

\begin{figure}
  \begin{center}
    \includegraphics[width=70mm]{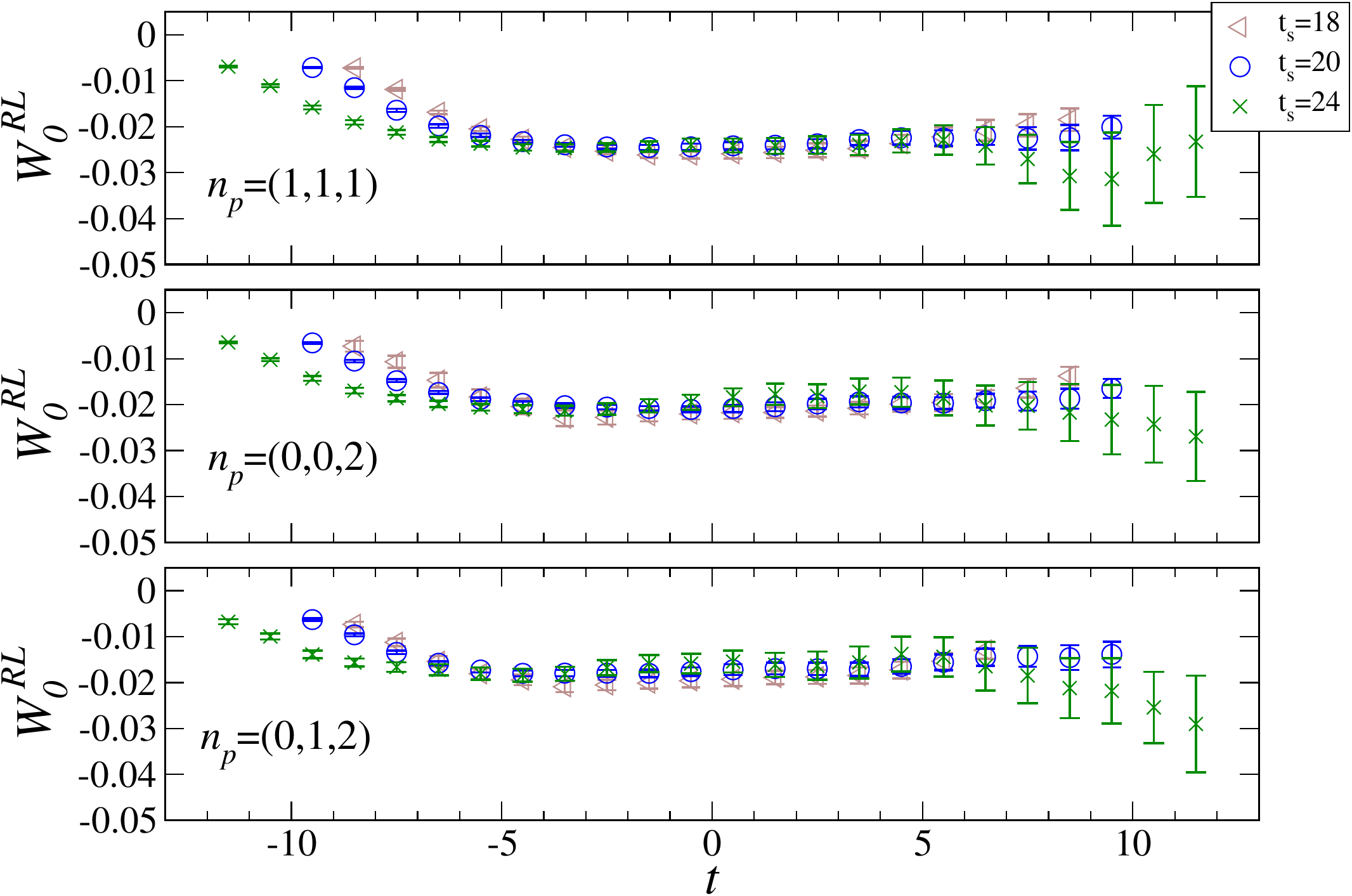}
    \hspace{3mm}
    \includegraphics[width=70mm]{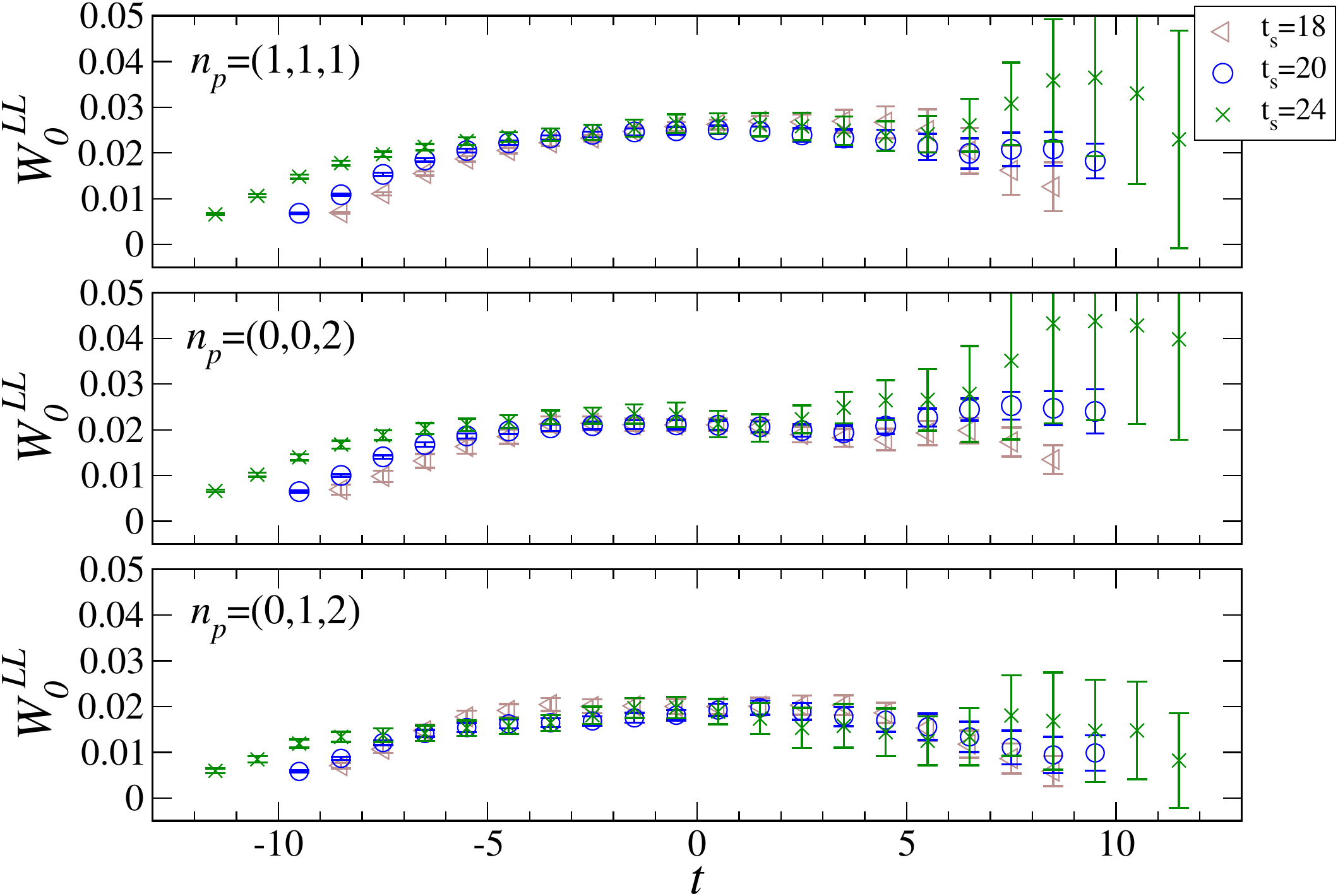}
    \caption{Time separation dependence of $W_0$ for $p\rightarrow\pi^0$ mode with Gaussian source in 139 MeV pion. (Left) $W_0^{RL}$, (Right) $W_0^{LL}$.
    The horizontal axis shows the shifted time-slice of the operator $t$
    such that $t=0$ always locates at the central position between the source and sink.
    }
    \label{fig:W0_tdep_pi}
  \end{center}
\end{figure}

We first show the dependence of source-sink separation for the $W_0$ of $p\rightarrow \pi_0$ channel with $RL$ operator to see the effect of the excited state contamination in Figure~\ref{fig:W0_tdep_pi}. This form factor is extracted from the combinations of the three-point function and two-point functions of nucleon and the pseudoscalar with certain projections (For detail, see e.g. Ref~\cite{Aoki:2006ib, Aoki:2007xm,Aoki:2017puj}.) One can see that an asymptotic plateau, which is independent of a choice of $t$, appears around the center of source-sink separation ($t=0$ in Figure~\ref{fig:W0_tdep_pi}) at three varieties of injection momenta even if $t_s$ is varied, and so that our smearing parameter can enough suppress the excited state contamination effect when $t_s\simeq 20$ corresponding to 1.7 fm separation.

\begin{figure}
  \begin{center}
    \includegraphics[width=72mm]{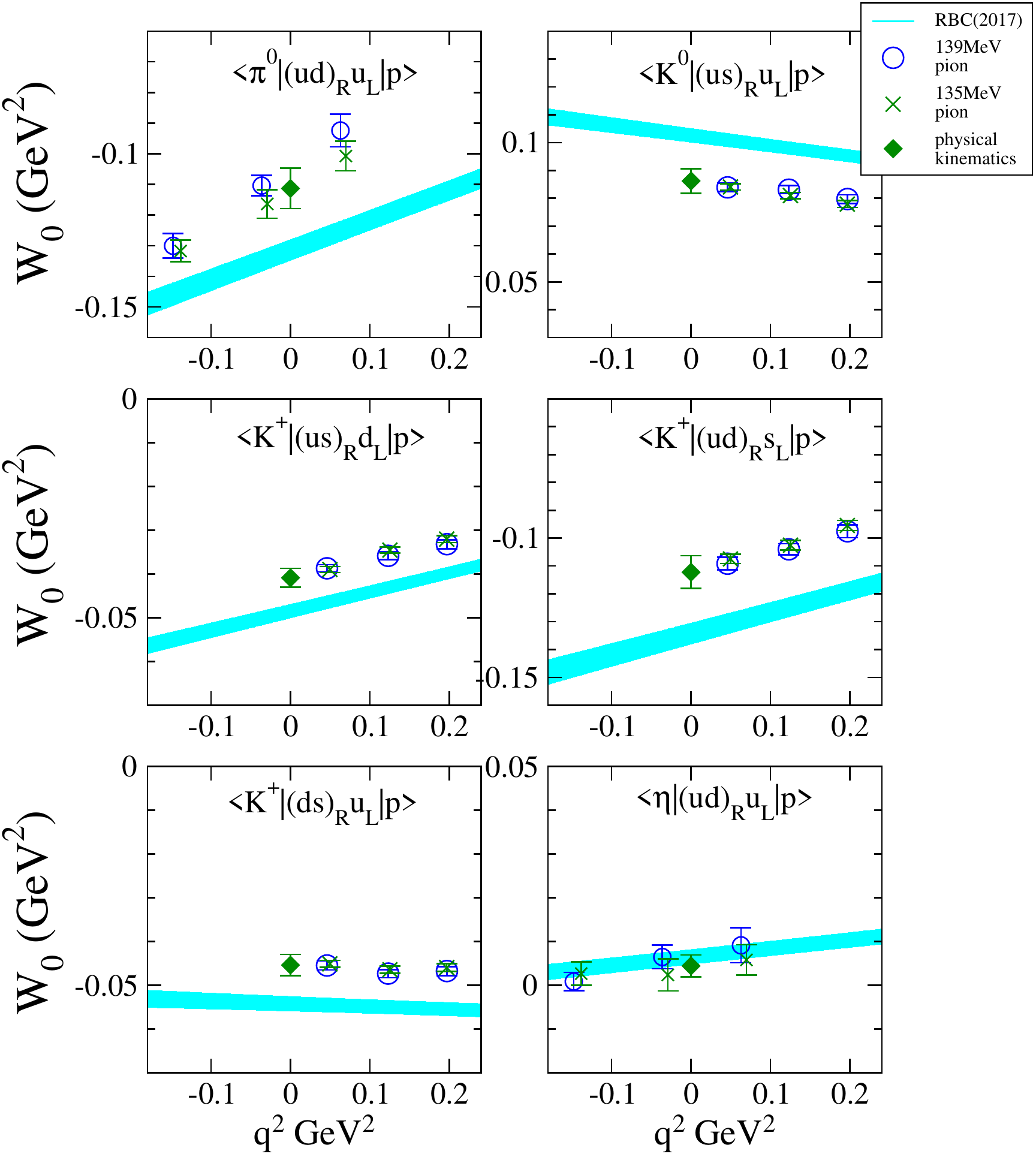}
    \hspace{3mm}
    \includegraphics[width=72mm]{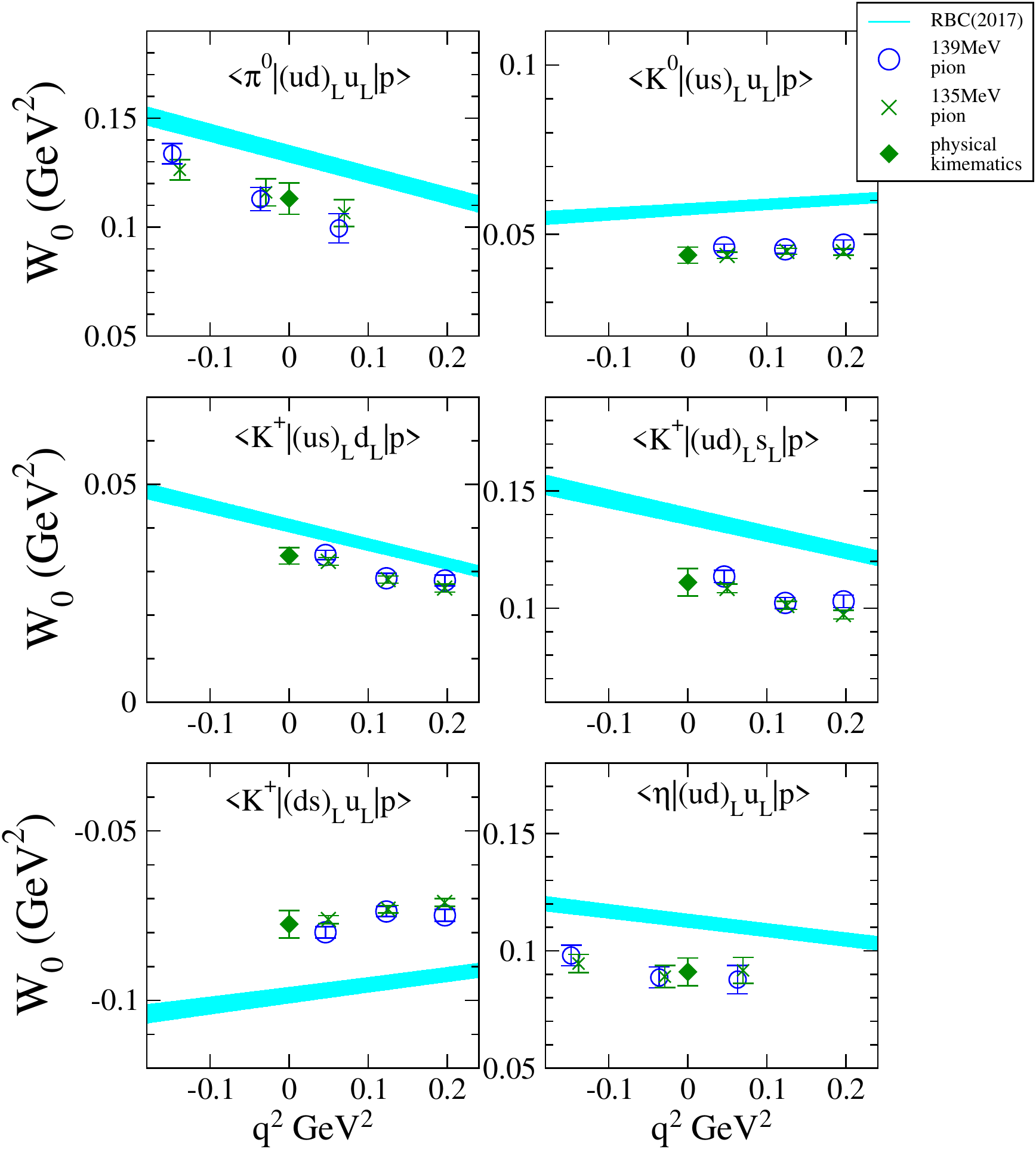}
    \caption{Transfer momentum dependence of $W_0^{RL}$ (Left) and $W_0^{LL}$ (Right) for 12 relevant transition modes with different masses in 139 MeV and 135 MeV pions. Colored bands represent the previous results of chiral extrapolation from the heavy pion mass by RBC~\cite{Aoki:2017puj}.}
    \label{fig:W0_q2dep}
  \end{center}
\end{figure}

Figure~\ref{fig:W0_q2dep} shows the transfer momentum dependence of the renormalized form factors $W_0^O(O=RL,LL)$ in 12 relevant transition modes. Taking an interpolation into the $q^2=-m_e^2\simeq0$ by fitting procedure, the values denoted as the physical kinematics in Fig.~\ref{fig:W0_q2dep} are obtained. Compared with the chiral extrapolated results of domain-wall fermions given from linear extrapolation used in the range of $340-700$ MeV~\cite{Aoki:2017puj} (colored band in Fig.~\ref{fig:W0_q2dep}), a discrepancy over 1$\sigma$ statistical error from our physical point calculation is seen. This discrepancy will be a possible effect of non-linear chiral behavior significantly appearing around physical point as suggested in the phenomenological model~\cite{Martin:2011nd}, however a detailed discussion will be followed only after the continuum extrapolation is taken.

%
%  FIG.3.1
%

\begin{figure}
  \begin{center}
    \includegraphics[width=80mm]{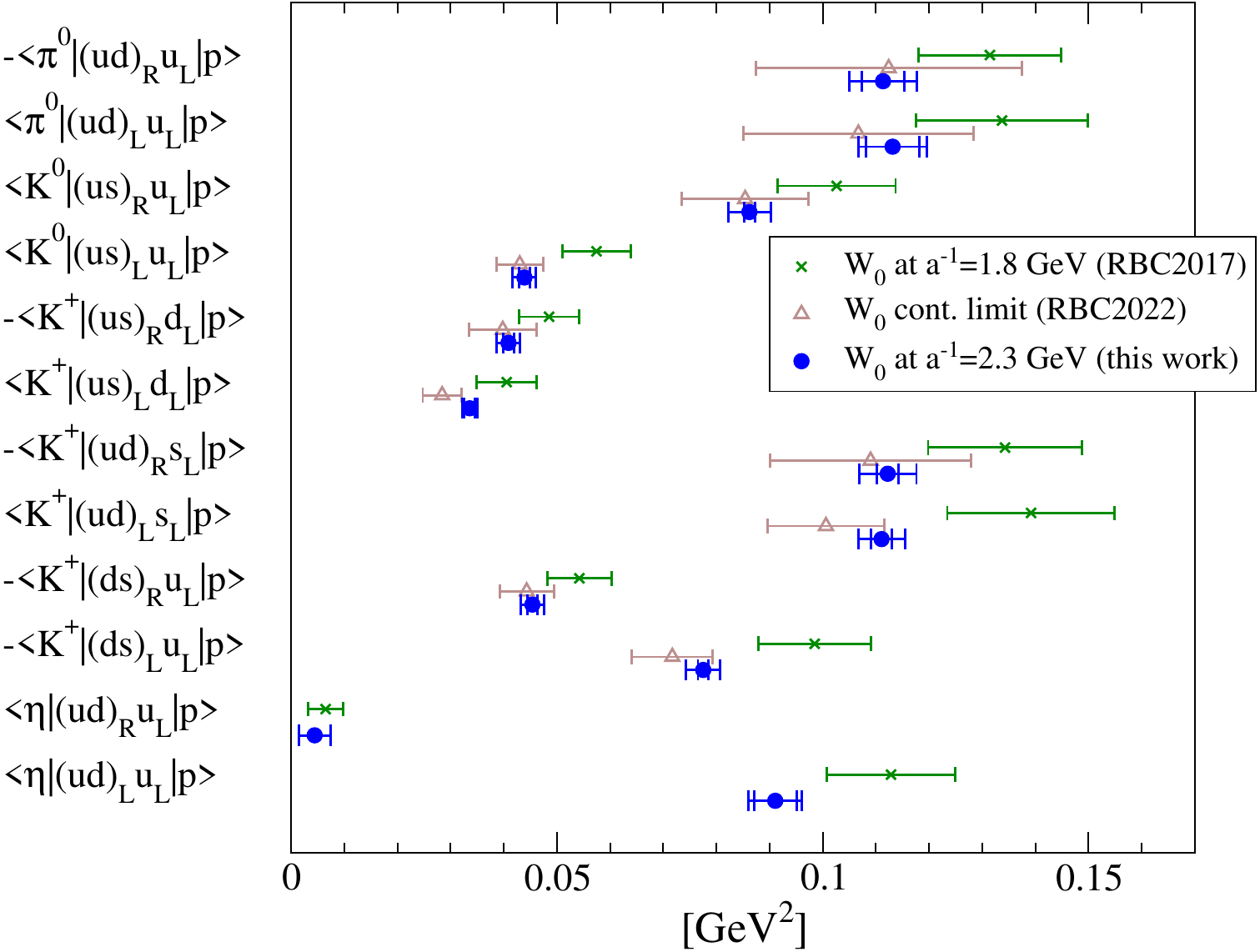}
    \caption{Summary of matrix elements of proton decay and comparison with other lattice results.}\label{fig:W0_summary}
  \end{center}
\end{figure}

The summary plot of our preliminary results of proton decay matrix elements after interpolation to the physical kinematics is shown in Figure~\ref{fig:W0_summary}. There are also the results in other LQCD collaboration~\cite{Yoo:2021gql}, and those are mostly consistent with ours. Note that the error bars of our result are given by the quadrature of the statistical error and the systematic error (renormalization constant, excited state contamination and operator mixing), however the discretization effect has not been accounted yet.

%
%  FIG.3.1
%

\section{Summary and Outlook}
\label{sec:summary_and_outlook}
We report our preliminary result of proton decay matrix elements in $64^4$ lattice volume with lattice $a = 0.085$ fm, at physical point for 2+1 flavor QCD. In this report, we update the computation of renormalization constant with two different intermediate schemes $\rm MOM_{3q}$ and $\rm SYM_{3q}$. After the systematic study of the excited-state contamination effect and the renormalization constant, our preliminary result is compatible with other LQCD result~\cite{Yoo:2021gql}. As in the near future work, the further studies of systematic uncertainties including the discretization effect are being planned.

\section*{Acknowledgement}
We would like to thank members of the PACS collaboration for useful discussions. Numerical calculations in this work were performed on Oakforest-PACS in Joint Center for Advanced High Performance Computing (JCAHPC) and Cygnus  and Pegasus in Center for Computational Sciences at University of Tsukuba under Multidisciplinary Cooperative Research Program of Center for Computational Sciences, University of Tsukuba, and Wisteria/BDEC-01 in the Information Technology Center, the University of Tokyo. This research also used computational resources of HPCI Project (Id: hp210143, hp240063). The  calculation employed OpenQCD system(http://luscher.web.cern.ch/luscher/openQCD/). This work is supported by the JLDG constructed over the SINET5 of NII. This work was also supported in part by Grants-in-Aid for Scientific Research from the Ministry of Education, Culture, Sports, Science and Technology (Nos. 21H00064, 23K03428) and MEXT as ``Program for Promoting Researches on the Supercomputer Fugaku'' (Search for physics beyond the standard model using large-scale lattice QCD simulation and development of AI technology toward next-generation lattice QCD; Grant Number JPMXP1020230409).

%--- bibliography ---------------------------------------------------  
\bibliographystyle{JHEP}
\bibliography{skeleton}

\end{document}